\documentstyle[11pt,aaspp4,flushrt]{article}

\tighten

\lefthead{Kriss et al.}
\righthead{HUT Far-Ultraviolet Observations of NGC 3516}

\begin{document}

\title{Far-Ultraviolet Observations of NGC 3516 using the Hopkins Ultraviolet
Telescope}

\author{G. A. Kriss, B. R. Espey, J. H. Krolik, Z. Tsvetanov,
W. Zheng, and A. F. Davidsen}
\affil{Department of Physics \& Astronomy, Johns Hopkins University,
    Baltimore, MD 21218}
\authoremail{gak@pha.jhu.edu}

\begin{abstract}
We observed the Seyfert 1 galaxy NGC 3516 twice
during the flight of Astro-2 using the Hopkins Ultraviolet Telescope
aboard the space shuttle {\it Endeavour} in March 1995.
Simultaneous X-ray observations were performed with {\it ASCA}.
Our far-ultraviolet spectra cover the spectral range
820--1840 \AA\ with a resolution of 2--4 \AA.
No significant variations were found between the two observations.
The total spectrum shows a red continuum, $f_\nu \sim \nu^{-1.89}$,
with an observed flux of
$\rm 2.2 \times 10^{-14}~erg~cm^{-2}~s^{-1}~\AA^{-1}$ at 1450 \AA,
slightly above the historical mean.
Intrinsic absorption in Lyman $\beta$ is visible as well as
absorption from
O~{\sc vi} $\lambda\lambda 1032,1038$,
N~{\sc v} $\lambda\lambda 1239,1243$,
Si~{\sc iv} $\lambda\lambda 1394,1403$, and
C~{\sc iv} $\lambda\lambda 1548,1551$.
The UV absorption lines are far weaker than is usual for NGC~3516, and
also lie closer to the emission line redshift rather than showing the
blueshift typical of these lines when they are strong.
The neutral hydrogen absorption, however, is blueshifted by
$400~\rm km~s^{-1}$ relative to the systemic velocity,
and it is opaque at the Lyman limit.
The sharpness of the cutoff indicates a low effective Doppler parameter,
$b < \rm 20~km~s^{-1}$.
For $b = \rm 10~km~s^{-1}$ the derived intrinsic column is
$\rm 3.5 \times 10^{17}~cm^{-2}$.
As in NGC~4151, a single warm absorber cannot produce the strong absorption
visible over the wide range of observed ionization states.
Matching both the UV and X-ray absorption simultaneously requires absorbers
spanning a range of $10^3$ in both ionization parameter and column density.
\end{abstract}


\section{Introduction}

Only 3--10\% of Seyfert 1 galaxies show intrinsic UV absorption in the
resonance lines of highly ionized elements (\cite{Ulrich88}).
Of these, NGC~3516 has shown the strongest and most variable absorption
lines (\cite{UB83}; \cite{Voit87}; \cite{Walter90}; \cite{Kolman93};
\cite{Koratkar96}).
NGC~3516 is unusual for a Seyfert 1 in other respects as well.
While as many as half of all Seyfert 1's show absorption
by ionized material intrinsic to the source, characterized as
a ``warm absorber" (\cite{NP94}),
the only Seyfert 1 besides NGC~3516 with equivalent neutral hydrogen columns
exceeding $5 \times 10^{22}~\rm cm^{-2}$ is NGC~4151 (\cite{Kolman93};
\cite{NP94}; \cite{Yaqoob89}; \cite{Yaqoob93}).
In NGC~4151, these variable columns range from 1 to
$12 \times 10^{22}~\rm cm^{-2}$ (\cite{Yaqoob89}; \cite{Yaqoob93}), and a
similar
range of variation has been observed in NGC~3516 (\cite{Kolman93}; \cite{NP94};
\cite{Kriss96b}).
Extended X-ray emission has been seen in the nuclei of Seyfert 2 galaxies
(\cite{Wilson92}; \cite{Weaver95}), but among Seyfert 1's the only examples are
NGC~4151 (\cite{EBH83}; \cite{Morse95}) and possibly NGC~3516 (\cite{Morse95}).
NGC~3516 is also one of the rare Seyfert 1's with an extended narrow-line
region (NLR) having a biconical morphology (\cite{UP80}; \cite{Pogge89};
\cite{Miyaji92}; \cite{Golev95}).
Again, NGC~4151 is the only similar counterpart (\cite{Evans93}).
This rare combination of strong UV and X-ray absorption and extended
narrow-line
emission suggests that they may be related phenomena.

Extended narrow-line emission in Seyfert galaxies is commonly associated
with photoionization by a collimated source of radiation.
Biconical morphologies are most often found in Seyfert 2 galaxies
(\cite{Pogge89}; \cite{Evans94}; \cite{SK96}),
and they suggest that our line of sight
to the central source of radiation is obscured.
In the context of unified models of Seyfert galaxies (see the review by
\cite{Antonucci93}), Seyfert 1's present us with a direct line of sight to
the broad emission-line region (BELR) and continuum source, whereas our line
of sight in Seyfert 2's is blocked by a torus opaque from the mid-infrared to
at least soft X-rays.
If the torus collimates the ionizing radiation, then biconical morphologies
should not be observed in Seyfert 1 galaxies.
Given this line of reasoning, Evans et al. (1993)\markcite{Evans93} proposed
that the UV-absorbing material on our line of sight in NGC~4151,
but not in the torus proper, might collimate the ionizing radiation.

Although the strength of the UV and X-ray absorption in both NGC~4151 and
NGC~3516 suggests that the two absorbing mechanisms are related,
it is not clear how.
Kolman et al.'s (1993)\markcite{Kolman93} simultaneous X-ray and UV
observations
of NGC~3516 were inconclusive due to a lack of variability.
Common UV and X-ray absorption at much weaker levels in some active galactic
nuclei has been successfully modeled with a single warm absorber
(\cite{Mathur94}; \cite{Mathur95}), but the wide range of ionization states
of the UV absorber in NGC~4151 is not compatible with the simplest
warm absorber models (\cite{Kriss95}).

The similarities of NGC~3516 and NGC~4151 in their UV and X-ray absorption and
in their biconical NLR's prompted us to explore the far-UV spectrum of
NGC~3516 shortward of 1200 \AA\ using the Hopkins Ultraviolet Telescope (HUT).
Our goal was to search for further evidence that UV and
X-ray absorbing gas in active galactic nuclei (AGN) is related to the
collimation mechanism for the ionizing radiation.
To improve our understanding of the relationship between the UV and
X-ray absorbing gas, we also performed simultaneous X-ray observations
using the Japanese X-ray satellite {\it ASCA}.
In this paper we present the far-UV spectrum obtained with HUT.
A companion paper (Kriss et al. 1996) discusses the {\it ASCA} observations.

\section{Observations}

\begin{figure}[t]
\plotfiddle{"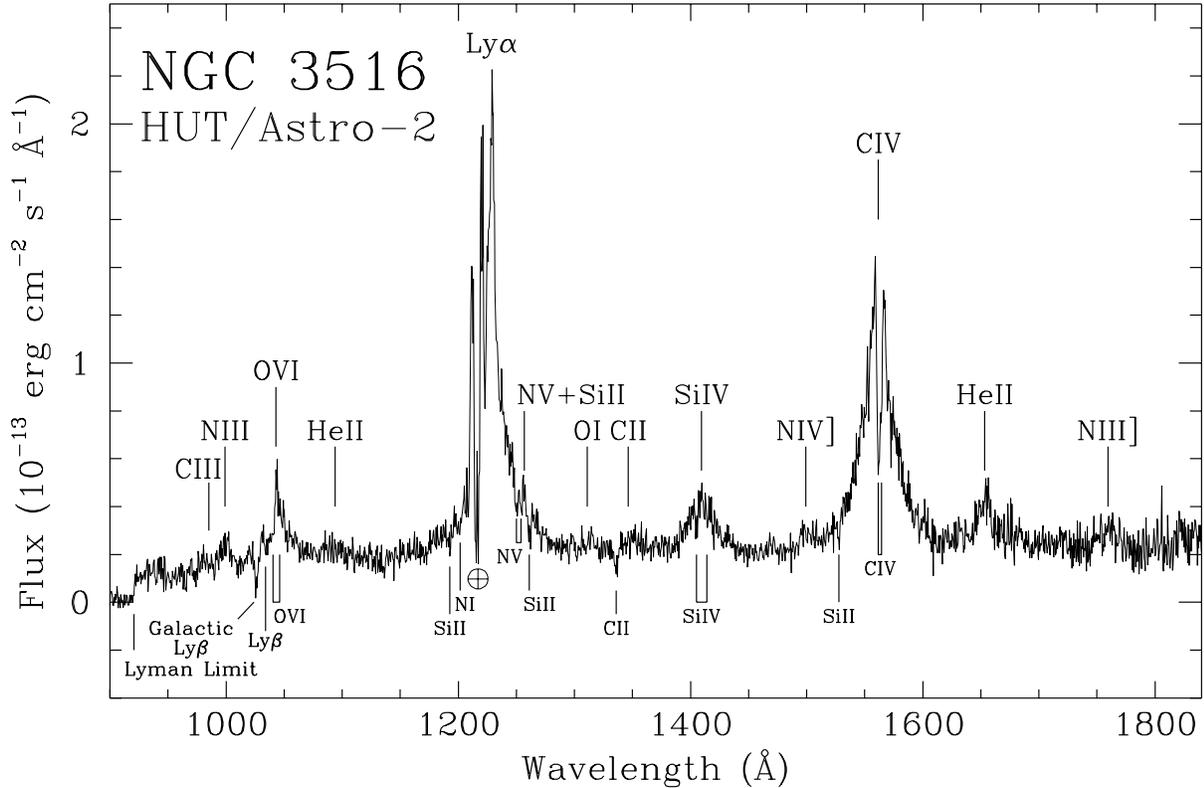"}{3.8in}{-90}{65}{65}{-265}{355}
\caption{
The flux-calibrated spectrum of NGC~3516 obtained with the Hopkins
Ultraviolet Telescope during the Astro-2 mission is shown.
Significant emission and absorption lines are marked.
The indicated Lyman limit is at a redshift of 0.0075, and it is
intrinsic to NGC~3516 as are
the absorption lines of {\sc O~vi}, {\sc N~v}, Si~{\sc iv}, and {\sc C~iv}.
The remaining absorption lines are likely galactic in origin.
The earth symbol indicates the residual feature produced by
subtraction of geocoronal Ly$\alpha$ emission.
\label{n3516uv1pp.ps}}
\end{figure}

We observed NGC~3516 on two occasions during the Astro-2 mission, once
for 1518~s beginning at 5:35:34 UT on 1995 March 11, and again for 2200~s
beginning at 1:32:03 on 1995 March 13.  Both observations were through a
20\arcsec\ aperture during orbital night when airglow is at a minimum.
The basic design of HUT is described by Davidsen et al.
(1992)\markcite{Davidsen92}.
Briefly, a 0.9-m mirror collects light for a prime-focus, Rowland-circle
spectrograph.  A photon-counting detector sensitive from 820--1840 \AA\
samples the dispersed spectrum at a resolution of 2--4 \AA\ with
$\sim$0.52 \AA\ per pixel.
Improvements to HUT, its performance during
the Astro-2 mission, and our basic data reduction procedures
are described by Kruk et al. (1995)\markcite{Kruk95}.
The raw data were reduced by subtracting
dark counts, correcting for scattered geocoronal Ly$\alpha$
emission and subtracting airglow. We then flux-calibrated the spectrum
using the time-dependent inverse sensitivity curves developed from
on-orbit observations and model atmospheres of white dwarfs.
Statistical errors for each pixel are calculated from the raw count spectra
assuming a Poisson distribution and are propagated through the data reduction
process.
As there was no evidence for variability in the UV data,
the two separate observations are weighted by their exposure times and combined
to form the mean flux-calibrated HUT spectrum of NGC~3516
shown in Figure \ref{n3516uv1pp.ps}.

To model the spectrum of NGC~3516 and measure properties of the continuum,
emission lines and the absorption lines, we use the IRAF\footnote{
The Image Reduction and Analysis Facility (IRAF) is distributed by
the National Optical Astronomy Observatories, which is operated by the
Association of Universities for Research in Astronomy, Inc. (AURA) under
cooperative agreement with the National Science Foundation.
}
task {\tt specfit} (\cite{Kriss94a}).
We fit the continuum with a power law in $f_\lambda$.
The brightest broad emission lines ({\sc O~vi} $\lambda 1034$, Ly$\alpha$, and
{\sc C~iv} $\lambda 1549$) are well described by power law profiles
while single Gaussian components are adequate for the weaker broad lines.
The power law profile has a functional form
$F_\lambda \propto ( \lambda / \lambda_o )^{\pm \alpha}$,
where $\alpha = {\rm ln}~2 / ( 1 + {\rm FWHM} / 2 c )$
(c.f. NGC~4151, \cite{Kriss92}).
Additional narrow Gaussian
cores are required for Ly$\alpha$ and He~{\sc ii} $\lambda1640$.
Single Gaussian profiles are used for all absorption lines
other than the Lyman series.
We allow extinction to vary freely following a Cardelli, Clayton, \& Mathis
(1989)\markcite{CCM89} curve with $\rm R_V = 3.1$.

To model hydrogen absorption in the Galaxy and in NGC~3516,
we compute grids of transmission functions including transitions up to
$\rm n = 50$.  Using Voigt profiles of varying column density and
Doppler parameter, we then convolve the transmission with the
instrument resolution.
Galactic neutral hydrogen is fixed at zero redshift at a column density
of $3.35 \times 10^{20}~\rm cm^{-2}$ (\cite{Stark92}) with a Doppler
parameter of $b = 10~\rm km~s^{-1}$.
A sharp, redshifted Lyman edge is readily apparent in the NGC~3516 spectrum.
Neutral hydrogen intrinsic to NGC~3516 is permitted to vary freely in
column density, redshift, and Doppler parameter.

\begin{figure}[t]
\plotfiddle{"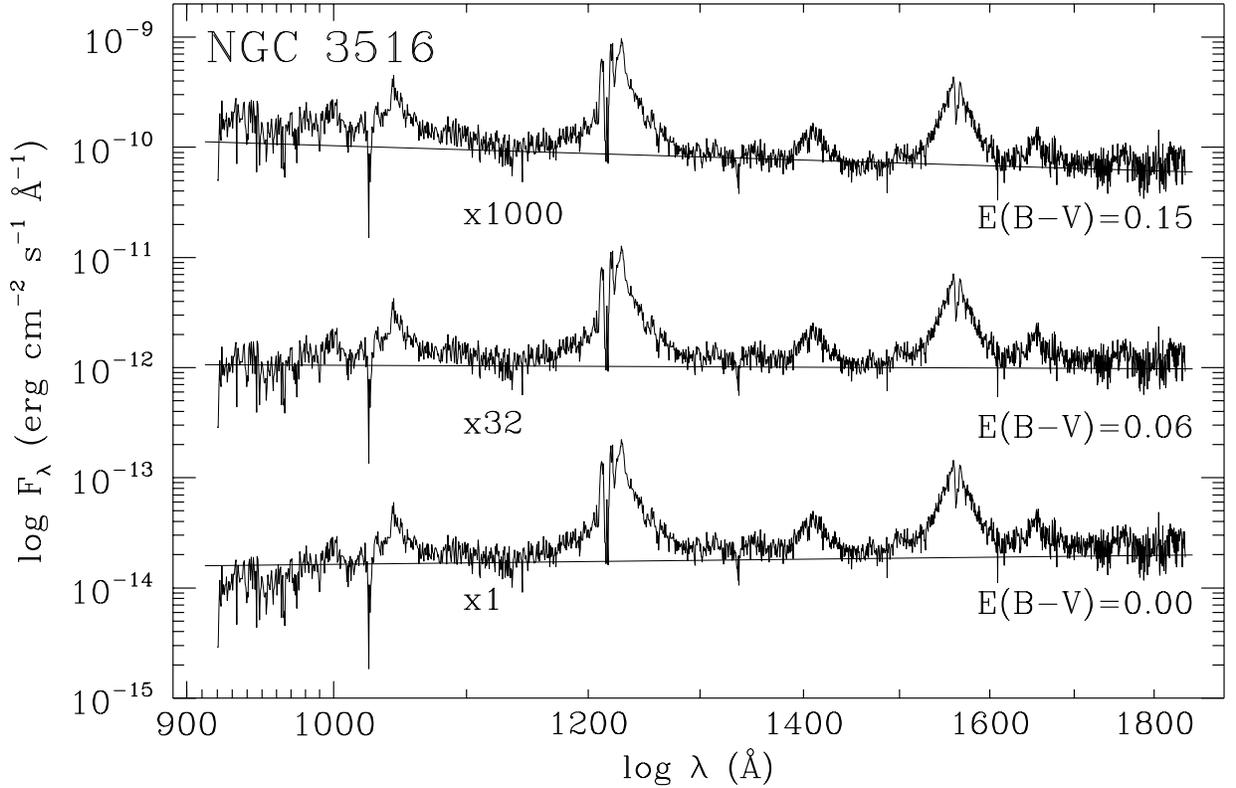"}{3.8in}{-90}{65}{65}{-255}{355}
\caption{
The figure shows the HUT spectrum of NGC~3516 on a log-log scale with
extinction corrections applied as indicated.
The spectra have been scaled up for display by the factors under each spectrum.
The solid straight lines through each spectrum are the best fit power law
continuum levels for the given extinction correction.
No correction clearly produces a deficit
in flux at short wavelengths, and $E(B - V) = 0.15$ leads to an excess in
continuum flux shortward of 1000 \AA.  $E(B - V) = 0.06$ provides the best fit.
Note that the broad wings on the emission lines leave clear continuum
visible only in narrow windows shortward of 980 \AA\ and near 1130 \AA,
1450 \AA\ and 1800 \AA.
\label{n3516uv2pp.ps}}
\end{figure}

Our best fit yields $\chi^2 / \nu = 1652 / 1582$
for 1671 data points between 916 \AA\ and 1800 \AA\
(we omit a region from 1207-1222 \AA\ surrounding geocoronal Ly$\alpha$).
The fitted continuum has
$f_\lambda~=~3.61~\times~10^{-14}~(\lambda / 1000 \rm \AA )^{-0.11}~
\rm erg~cm^{-2}~s^{-1}~\AA^{-1}$
with $E(B - V) = 0.06 \pm 0.01$.
In frequency space this corresponds to a
spectral index $\alpha = 1.89$ for $f_\nu \sim \nu^{- \alpha}$.
This is rather steep, but it is not that unusual for a low-luminosity AGN.
For comparison, the Astro-1 spectrum of NGC~4151 showed a spectral index of
1.50 (\cite{Kriss92}), the FOS spectrum of NGC~1566 has $\alpha = 1.53$
(\cite{Kriss91}), and the UV continuum of M~81 has $\alpha = 1.5 - 2.0$
(\cite{HFS96}).
Our fitted extinction is higher than the predicted Galactic reddening of
$E(B - V) = 0.02$--0.03 in the maps of Burstein \& Heiles
(1982)\markcite{BH82},
but compatible with that expected using
$N_{H I} = 3.35 \times 10^{20}~\rm cm^{-2}$ and a gas-to-dust ratio of
$N_{H I} / E(B - V) = 5.2 \times 10^{21}~\rm cm^{-2}$ (\cite{SVS85}).
Previous work based on IUE observations (\cite{Kolman93}; \cite{Koratkar96})
have determined $E(B - V) = 0.15$ based on the strength of the 2200 \AA\
absorption feature, but such a large extinction correction provides a poor
match to our data --- for $E(B - V) = 0.15$, $\chi^2 = 1782$.
To show the sensitivity of the data to the extinction correction,
Figure \ref{n3516uv2pp.ps} compares the observed spectrum with power-law
continua after correction for $E ( B - V ) = 0.06$ and 0.15.
The figure clearly shows that no extinction correction leads to a deficit in
continuum flux below 1000 \AA, whereas an extinction correction of
$E(B - V) = 0.15$ produces an excess in flux at short wavelengths.
We suggest that the apparent strength
of the 2200 \AA\ dip in the IUE data is not due to extinction, but rather
to the onset of broad emission from blended Fe~{\sc ii} at 2300 \AA.

\begin{figure}[t]
\plotfiddle{"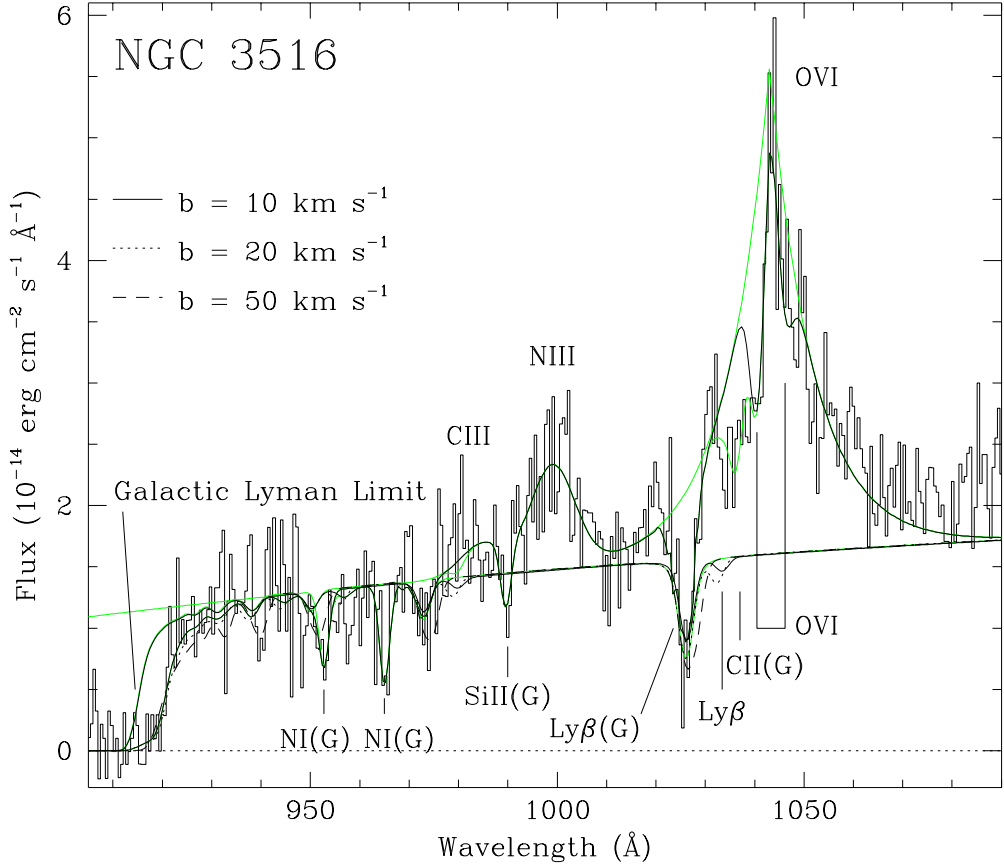"}{4.3in}{-90}{65}{65}{-265}{355}
\caption{
This expanded view of the short-wavelength end of the HUT spectrum shows the
intrinsic Lyman limit and the {\sc O~vi} absorption more clearly.
The position of the Galactic Lyman limit is shown to illustrate the intrinsic
absorption.  The solid, dotted, and dashed curves show the shape of the
intrinsic Lyman absorption in NGC~3516 for assumed Doppler parameters of
$b = 10$, 20, and 50 $\rm km~s^{-1}$, respectively.  As described in the text,
at 90\% confidence, $b < 20~\rm km~s^{-1}$.
The heavy and light solid lines show our fit to the {\sc O~vi} emission and
absorption.  The light solid line shows the {\sc O~vi} emission profile
superposed on the powerlaw continuum.  The heavy solid line shows the
contribution of the intrinsic {\sc O~vi} absorption to the fit.
The light solid line in the vicinity of 1037 \AA\ shows the contribution
of intrinsic Ly$\beta$ and Galactic {\sc C~ii} $\lambda 1037$ to the fit.
Galactic absorption features in the spectrum are denoted by a ``(G)"
following the line identification.
\label{n3516uv3pp.ps}}
\end{figure}

For the neutral hydrogen intrinsic to NGC~3516 we find a best fit redshift
of $\rm z = 0.0075 \pm 0.0005$, $\rm N_{HI} = 3.5 \times 10^{17}~cm^{-2}$, and
$b = 10~\rm km~s^{-1}$.
This is blue-shifted by $400 \pm 150~\rm km~s^{-1}$ relative to the
$2649~\rm km~s^{-1}$ systemic velocity of NGC~3516 measured using the stellar
absorption lines (\cite{VC85}).
The opaque Lyman limit requires neutral hydrogen with a minimum column
density $\rm N_{HI} > 2.2 \times 10^{17}~cm^{-2}$ (90\% confidence).
Its sharpness limits $b$ to less than $20~\rm km~s^{-1}$ at 90\% confidence.
The effect of the assumed Doppler parameter on the shape of the intrinsic
Lyman limit is shown with an enlarged view of the Lyman-limit region in
NGC~3516 in Figure \ref{n3516uv3pp.ps}.
The general weakness of the Lyman absorption lines (only Ly$\alpha$ and
perhaps Ly$\beta$ are detected)
gives an upper limit of $\rm N_{HI} < 6.3 \times 10^{17}~cm^{-2}$
at 90\% confidence.

Measured properties of the fitted emission lines are in Table
\ref{tab-hutflux}.
The fitted absorption lines are summarized in Table \ref{tab-hutabs}.
All tabulated features, except for Ly$\beta$,
have a statistical significance exceeding $3\sigma$.
The error matrix of the fit is used to derive the $1\sigma$ statistical errors
shown in the table.
Systematic errors of up to $\sim$5\% are likely to be present in the fluxes,
and our wavelength scale has a limiting accuracy of $\sim 60~\rm km~s^{-1}$.
For ease in fitting, some poorly constrained parameters were linked to others
with better determined values.
For example, the velocity offset and FWHM of the {\sc C~iii} $\lambda 977$
and the {\sc N~iii} $\lambda 991$ lines were linked to share common values,
and the FWHM of He~{\sc ii} $\lambda 1085$ was set equal to that of
He~{\sc ii} $\lambda 1640$.
Any entries in the tables with identical velocity offsets or FWHM's were
linked in a similar manner.
Since the intrinsic {\sc O~vi} absorption is difficult to see in
Figure \ref{n3516uv1pp.ps}, an enlarged view illustrating our fits to this
region is shown in Figure \ref{n3516uv3pp.ps}.
The {\sc O~vi} absorption doublet straddles the peak of the {\sc O~vi} emission
line, and the sharpness of the line peak is largely defined by the flux
escaping at wavelengths between the doublets.
Galactic {\sc C~ii}, Galactic Ly$\beta$, and intrinsic Ly$\beta$ absorption
all contribute to the deficit in flux on the blue wing
of the {\sc O~vi} emission.

\begin{deluxetable}{ l c c c c }
\tablecolumns{5}
\tablewidth{440pt}
\tablecaption{Emission Lines in the HUT Spectrum of NGC~3516
\label{tab-hutflux}}
\tablehead{
\colhead{Line} & \colhead{$\rm \lambda_{vac}$} &
\colhead{Flux\tablenotemark{a}}
 & \colhead{Velocity\tablenotemark{b}} & \colhead{FWHM} \\
\colhead{} & \colhead{(\AA)} & \colhead{} & \colhead
{$\rm ( km~s^{-1} )$} & \colhead{$\rm ( km~s^{-1} )$} \\
}
\startdata
C~III      & \phantom{0}977.02 & $\phantom{0}0.7 \pm 0.3$ & \phantom{0}$-235
\pm \phantom{0}249$ & $\phantom{0}3190 \pm \phantom{0}469$ \nl
N~III      & \phantom{0}990.83 & $\phantom{0}2.3 \pm 0.3$ & \phantom{0}$-235
\pm \phantom{0}249$ & $\phantom{0}3190 \pm \phantom{0}469$ \nl
O~VI       & 1033.83 & $15.9 \pm 1.1$ & $\phantom{-00}13 \pm \phantom{00}63$ &
$\phantom{0}3516 \pm \phantom{0}375$ \nl
He~II      & 1085.15 & $\phantom{0}2.9 \pm 0.6$ & $-2012 \pm \phantom{0}880$ &
$10041 \pm 1694$ \nl
Ly$\alpha$ & 1215.67 & $\phantom{0}5.6 \pm 0.9$ & $\phantom{-0}721 \pm
\phantom{00}33$ & $\phantom{00}808 \pm \phantom{00}98$ \nl
Ly$\alpha$ & 1215.67 & $84.3 \pm 3.1$ & \phantom{0}$-345 \pm \phantom{0}160$ &
$\phantom{0}4556 \pm \phantom{0}230$ \nl
Ly$\alpha$ & 1215.67 & $89.9 \pm 3.2$ & \nodata & \nodata \nl
N~V        & 1240.15 & $\phantom{0}5.2 \pm 1.0$ & $\phantom{-}1443 \pm
\phantom{0}339$ & $\phantom{0}5660 \pm \phantom{0}181$ \nl
O~I        & 1304.35 & $\phantom{0}1.3 \pm 0.3$ & $-1152 \pm \phantom{0}573$ &
$\phantom{0}5660 \pm \phantom{0}181$ \nl
C~II       & 1335.30 & $\phantom{0}1.9 \pm 0.3$ & $\phantom{-0}285 \pm
\phantom{0}103$ & $\phantom{0}5660 \pm \phantom{0}181$ \nl
Si~IV      & 1393.76 & $\phantom{0}6.1 \pm 0.2$ & $\phantom{-0}285 \pm
\phantom{0}103$ & $\phantom{0}5660 \pm \phantom{0}181$ \nl
Si~IV      & 1402.77 & $\phantom{0}3.1 \pm 0.1$ & $\phantom{-0}285 \pm
\phantom{0}103$ & $\phantom{0}5660 \pm \phantom{0}181$ \nl
N~IV]      & 1486.50 & $\phantom{0}1.1 \pm 0.2$ & \phantom{00}$-49 \pm
\phantom{0}242$ & $\phantom{0}2152 \pm \phantom{0}498$ \nl
C~IV       & 1549.05 & $58.4 \pm 0.6$ & \phantom{0}$-191 \pm \phantom{00}65$ &
$\phantom{0}3705 \pm \phantom{0}117$ \nl
He~II      & 1640.50 & $\phantom{0}2.4 \pm 0.5$ & \phantom{0}$-280 \pm
\phantom{0}113$ & $\phantom{0}1869 \pm \phantom{0}285$ \nl
He~II      & 1640.50 & $\phantom{0}5.7 \pm 0.9$ & \phantom{0}$-280 \pm
\phantom{0}113$ & $10041 \pm 1694$ \nl
He~II      & 1640.50 & $\phantom{0}8.2 \pm 1.0$ & \nodata & \nodata \nl
N~III]     & 1750.51 & $\phantom{0}2.2 \pm 0.4$ & $-1084 \pm \phantom{0}238$ &
$\phantom{0}2870 \pm \phantom{0}476$ \nl
\enddata
\tablenotetext{a}{
Flux in $\rm 10^{-13}~erg~cm^{-2}~s^{-1}$ corrected for $E(B - V) = 0.06$.
}
\tablenotetext{b}{
Velocity relative to a systemic redshift of $\rm cz = 2649~km~s^{-1}$
(\cite{VC85}).
}
\end{deluxetable}

\begin{deluxetable}{ l c c c c }
\tablecaption{Absorption Lines in the HUT Spectrum of NGC 3516
\label{tab-hutabs}}
\tablecolumns{5}
\tablewidth{440pt}
\tablehead{
\colhead{Line} & \colhead{$\rm \lambda_{vac}$} & \colhead{EW} &
\colhead{Velocity\tablenotemark{a}} & \colhead{FWHM} \\
\colhead{} & \colhead{(\AA)} & \colhead{(\AA)} & \colhead{$\rm ( km~s^{-1} )$}
& \colhead{$\rm ( km~s^{-1} )$} }
\startdata
N~I        & \phantom{0}953.77 & $\phantom{0}1.3 \pm 0.4$ & $-2944 \pm
\phantom{0}146$ & $\phantom{00}804 \pm \phantom{0}274$ \nl
N~I        & \phantom{0}964.24 & $\phantom{0}1.6 \pm 0.5$ & $-2392 \pm
\phantom{0}115$ & $\phantom{00}804 \pm \phantom{0}274$ \nl
C~III      & \phantom{0}977.02 & $\phantom{0.0}< 1.0 \phantom{\pm 0.0}$ &
$\phantom{-000}0 \phantom{ \pm 0000}$ & $\phantom{00}800\tablenotemark{b}
\phantom{ \pm 0000}$ \nl
Si~II      & \phantom{0}989.87 & $\phantom{0}0.8 \pm 0.3$ & $-2704 \pm
\phantom{0}107$ & $\phantom{00}804 \pm \phantom{0}274$ \nl
Ly$\beta$ & 1025.72 & $\phantom{0}2.1 \pm 0.4$ & $-2649 \pm \phantom{0}156$ &
$\phantom{00}880\tablenotemark{b} \phantom{ \pm 0000}$ \nl
Ly$\beta$ & 1025.72 & $\phantom{0}0.2 \pm 0.4$ & \phantom{0}$-400 \pm
\phantom{0}150$ & $\phantom{00}880\tablenotemark{b} \phantom{ \pm 0000}$ \nl
C~II       & 1036.90 & $\phantom{0}1.2 \pm 0.3$ & $-2798 \pm \phantom{0}109$ &
$\phantom{0}1076 \pm \phantom{0}146$ \nl
O~VI       & 1031.93 & $\phantom{0}1.6 \pm 0.4$ & \phantom{0}$-161 \pm
\phantom{00}81$ & $\phantom{0}1076 \pm \phantom{0}146$ \nl
O~VI       & 1037.62 & $\phantom{0}0.7 \pm 0.4$ & \phantom{0}$-161 \pm
\phantom{00}81$ & $\phantom{0}1076 \pm \phantom{0}146$ \nl
Si~II      & 1193.14 & $\phantom{0}2.2 \pm 0.4$ & $-2733 \pm \phantom{0}127$ &
$\phantom{0}1645 \pm \phantom{0}255$ \nl
N~I        & 1200.16 & $\phantom{0}2.6 \pm 0.5$ & $-2269 \pm \phantom{0}122$ &
$\phantom{0}1645 \pm \phantom{0}255$ \nl
Ly$\alpha$ & 1215.67 & $\phantom{0}0.5 \pm 0.2$ & \phantom{0}$-400 \pm
\phantom{0}150$ & $\phantom{00}770\tablenotemark{b} \phantom{ \pm 0000}$ \nl
N~V        & 1238.82 & $\phantom{0}0.9 \pm 0.2$ & $\phantom{-00}63 \pm
\phantom{00}53$ & $\phantom{00}632 \pm \phantom{0}116$ \nl
N~V        & 1242.80 & $\phantom{0}0.4 \pm 0.2$ & $\phantom{-00}63 \pm
\phantom{00}53$ & $\phantom{00}632 \pm \phantom{0}116$ \nl
Si~II      & 1260.42 & $\phantom{0}1.3 \pm 0.3$ & $-2456 \pm \phantom{00}63$ &
$\phantom{00}831 \pm \phantom{0}170$ \nl
C~II       & 1335.30 & $\phantom{0}1.6 \pm 0.3$ & $-2454 \pm \phantom{00}58$ &
$\phantom{00}789 \pm \phantom{0}146$ \nl
Si~IV      & 1393.76 & $\phantom{0}0.8 \pm 0.4$ & \phantom{0}$-204 \pm
\phantom{00}75$ & $\phantom{00}778 \pm \phantom{0}585$ \nl
Si~IV      & 1402.77 & $\phantom{0}0.4 \pm 0.3$ & \phantom{0}$-204 \pm
\phantom{00}75$ & $\phantom{00}778 \pm \phantom{0}585$ \nl
Si~II      & 1527.17 & $\phantom{0}0.6 \pm 0.2$ & $-2495 \pm \phantom{00}47$ &
$\phantom{00}324 \pm \phantom{0}106$ \nl
C~IV       & 1548.19 & $\phantom{0}1.9 \pm 0.1$ & \phantom{00}$-15 \pm
\phantom{00}22$ & $\phantom{00}516 \pm \phantom{00}37$ \nl
C~IV       & 1550.77 & $\phantom{0}1.0 \pm 0.1$ & \phantom{00}$-15 \pm
\phantom{00}22$ & $\phantom{00}516 \pm \phantom{00}37$ \nl
\enddata
\tablenotetext{a}{
Velocity relative to a systemic redshift of $\rm cz = 2649~km~s^{-1}$
(\cite{VC85}).
}
\tablenotetext{b}{
FWHM fixed at the instrument resolution at this wavelength.
}
\end{deluxetable}

\section{The Broad Emission Lines}

The broad emission lines in NGC~3516 as observed with HUT show subtle
differences compared to other Seyfert 1's and low redshift AGN.
The optical to X-ray spectral index for NGC~3516, $\alpha_{ox} = 1.20$,
is typical of other Seyfert 1's (\cite{KC85}),
but the flux ratio of {\sc O~vi} $\lambda 1034$
to Ly$\alpha$ is only 0.18.  This lies below the correlation
of {\sc O~vi}/Ly$\alpha$ with $\alpha_{ox}$ discussed by Zheng, Kriss, \&
Davidsen (1995)\markcite{ZKD95}, but is comparable to the
mean value of 0.17 seen in high redshift quasars (\cite{Laor94}).
As in the case of Fairall~9, however, the lack of a soft X-ray excess in
NGC~3516 may be a significant factor in producing lower {\sc O~vi}/Ly$\alpha$
compared to other Seyfert 1's (\cite{Zheng95}).

The other noticeable difference is the relatively high strength of broad
{\sc C~iii} $\lambda 977$ and {\sc N~iii} $\lambda 991$ emission
in our spectrum.
Although these lines have been seen in other low-redshift quasars
(\cite{Laor94}; \cite{Laor95}), they are not detected in other HUT spectra of
Seyfert 1's.  The ratios of these lines to each other and to {\sc O~vi}
$\lambda 1034$ is also typical of that seen in the low redshift
quasars observed with HST (\cite{Laor95}).

Reverberation mapping experiments have shown that the broad emission-line
region (BELR) of AGN is highly stratified in both spatial distribution
and in ionization parameter (e.g., \cite{Clavel91}; \cite{Krolik91};
\cite{Reichert94}; \cite{Korista95}).
Nevertheless, it is useful to match a single-zone photoionization model to
the observed broad line emission in NGC~3516 to obtain a fiducial
for comparison to other Seyferts.  In addition, it provides a comparison to
models of the warm absorbing medium seen in the X-ray spectrum and in the
UV absorption lines that we discuss in \S\,4.
We use the photoionization code XSTAR (\cite{KK93}) to compute a grid of
photoionization models varying the total column density $N$ and the ionization
parameter $U = n_{ion} / n_H$, where
$n_{ion}$ is the number density of ionizing
photons between 13.6 eV and 13.6 keV illuminating the cloud and
$n_H$ is the density of hydrogen atoms.
As we are constraining only high ionization lines that are insensitive to the
density, we assume constant density clouds with $n_H = 10^9~\rm cm^{-3}$ and
solar abundances.
For the incident photoionizing continuum we use the extinction-corrected UV and
absorption-corrected X-ray spectrum of NGC~3516 as
described by Kriss et al. (1996)\markcite{Kriss96b}.
The UV power law was extrapolated to higher energies following
$f_{\nu} \sim \nu^{-1.89}$ with a break at 51 eV to the slope of the X-ray
power law, $f_{\nu} \sim \nu^{-0.78}$.
To match the observed line ratios we varied the parameters until we achieved
the closest fit to the strongest lines --- {\sc O~vi}+Ly$\beta$, Ly$\alpha$,
Si~{\sc iv}+{\sc O~iv}], {\sc C~iv}, and He~{\sc ii} $\lambda 1640$.
The closest match is for an ionization parameter $U = 0.035$.
The resulting line ratios from this model are compared to the observed values
and their error bars in Table \ref{tab-photo}.
Choosing $U = 0.035$ is mainly a balance between
the {\sc O~vi} emission and the {\sc C~iv} line.
Producing sufficient {\sc O~vi} requires $U = 0.045$, but then
the {\sc C~iv}/Ly$\alpha$ ratio becomes too high.
$U = 0.030$ fits the {\sc C~iv}/Ly$\alpha$ intensity ratio well, but the
{\sc O~vi} emission is then a factor of 2 less than observed.
As the lines we are matching originate in the high ionization illuminated
faces of the BELR clouds, their intensities are rather insensitive to changes
in the column density above $\log N = 21.6$ in our models.

\begin{deluxetable}{ l c c }
\tablecolumns{3}
\tablewidth{440pt}
\tablecaption{Observed and Modeled Emission Line Ratios in
NGC~3516\label{tab-photo}}
\tablehead{
\colhead{Feature} & \colhead{$I / I_{\lambda 1216}$\tablenotemark{a}} &
Model\tablenotemark{b}}
\startdata
{\sc C~iii} $\lambda 977$ & $0.008 \pm 0.003$ & 0.002 \nl
{\sc N~iii} $\lambda 991$ & $0.026 \pm 0.004$ & 0.000 \nl
{\sc O~vi}+Ly$\beta$ & $0.177 \pm 0.014$ & 0.123 \nl
He~{\sc ii} $\lambda 1085$ & $0.032 \pm 0.007$ & 0.009 \nl
{\sc N~v} $\lambda 1240$ & $0.058 \pm 0.011$ & 0.064 \nl
{\sc O~i} $\lambda 1304$ & $0.014 \pm 0.003$ & 0.000 \nl
{\sc C~ii} $\lambda 1335$ & $0.021 \pm 0.003$ & 0.007 \nl
Si~{\sc iv}+{\sc O~iv} $\lambda 1400$ & $0.102 \pm 0.005$ & 0.102 \nl
{\sc N~iv}] $\lambda 1486$ & $0.012 \pm 0.002$ & 0.074 \nl
{\sc C~iv} $\lambda 1549$ & $0.650 \pm 0.024$ & 0.778 \nl
He~{\sc ii} $\lambda 1640$ & $0.091 \pm 0.012$ & 0.037 \nl
{\sc N~iii}] $\lambda 1750$ & $0.024 \pm 0.004$ & 0.012 \nl
\enddata
\tablenotetext{a}{
Line ratios are relative to the major contributors to the $\lambda 1216$
blend --- H Ly$\alpha$, He~{\sc ii} $\lambda 1216$,
and {\sc O~v} $\lambda 1218$.}
\tablenotetext{b}{
Line ratio computed for a model with $U = 0.035$, $n_H = 10^9~\rm cm^{-3}$,
and $N = 10^{22}~\rm cm^{-2}$.
}
\end{deluxetable}

The best-fit ionization parameter $U = 0.035$ is a typical value for
single-zone models of AGN broad-line regions, but one can see from
Table \ref{tab-photo} that this model is insufficient to explain all the
observed line ratios.
The most noticeable differences are in the relatively high strengths observed
for {\sc C~iii} $\lambda 977$, {\sc N~iii} $\lambda 991$, and
He~{\sc ii} $\lambda 1640$.
Given the compromise we made between matching {\sc O~vi} and {\sc C~iv},
one can envision that a higher ionization zone producing relatively more
{\sc O~vi} and He~{\sc ii} $\lambda 1640$ could account for part of these
differences.
An additional population of higher density clouds may be required to
produce the enhanced {\sc C~iii} $\lambda 977$, which becomes a more
important coolant as other lines become optically thick (\cite{Netzer90}).
The strong {\sc N~iii} $\lambda 991$ is more of a puzzle.
Most photoionization models predict it to be weaker than
{\sc C~iii} $\lambda 977$, yet in our spectrum and in the HST quasar spectra
(\cite{Laor95}) it is stronger.
It is possible that fluorescent mechanisms could enhance the strength of
this transition under favorable circumstances (\cite{FFP95}).

Although the broad lines in NGC~3516 seem to be mostly produced in a region
with an ionization parameter typical of the BELR in other AGN,
$U = 0.035$ is an order of magnitude lower than the ionization parameter
required for the warm absorbing gas detected in the ASCA X-ray spectrum
(\cite{Kriss96b}).
This does not rule out an origin in the BELR for the X-ray warm absorbers,
but these absorbers must be physically distinct from the clouds producing
the bulk of the broad-line emission.

\section{The Complex Absorbing Medium in NGC 3516}

Intrinsic UV absorption lines are prominent in our spectrum of NGC~3516, but
they are near the weakest levels seen in NGC~3516, comparable to their
appearance in the 1993 IUE monitoring campaign (\cite{Koratkar96}).
The frequently observed blue-shifted absorption trough (\cite{Voit87};
\cite{Walter90}; \cite{Kolman93}) is not present, and the {\sc C~iv} feature
in our spectrum probably corresponds to the narrow component in
the models of Walter et al. (1990)\markcite{Walter90}.
Although the {\sc C~iv} absorption line in the HUT spectrum is narrow, it is
resolved.  After correcting for instrumental broadening by subtracting
a 2.0 \AA\ Gaussian in quadrature (\cite{Kruk95}),
we find an intrinsic width for the line of 350 $\rm km~s^{-1}$.
This corresponds to $b = 210~\rm km~s^{-1}$, which is
consistent with the observed equivalent widths (EWs)
and optically thin doublet ratio of the deblended {\sc C~iv} lines.
The other high ionization resonance lines also have optically thin doublet
ratios consistent with $b \sim 200~\rm km~s^{-1}$.
In contrast, the neutral hydrogen absorption at the Lyman limit requires
$b < 20~\rm km~s^{-1}$.
This, and the significantly different velocity of the neutral hydrogen
absorption, are indications that the absorbing medium contains multiple zones.

The simultaneous X-ray observations of NGC~3516 discussed in the preceding
paper (\cite{Kriss96b}) also hint that the absorbing medium may be complex.
The ASCA observations show {\sc O~vii} and {\sc O~viii} edges
characteristic of warm absorbers that have been seen in other AGN
(\cite{Turner93}; \cite{Mathur94}; \cite{Fabian94a}; \cite{Fabian94b}).
Photoionized warm absorber models similar to those discussed by
Krolik \& Kriss (1995)\markcite{KK95} can fit the data,
but the fit is significantly improved if two absorbing zones with differing
ionization parameter and total column density are used.
These warm absorber models use the same
incident photoionizing continuum described in the previous section.
We also considered alternative ionizing continua to test the sensitivity of
our results to the assumed incident spectrum.
We used both an extremely hard spectrum with $f_\nu \sim \nu^{-0.78}$ from
2500 \AA\ through the UV and X-ray up to 100 keV,
and also the spectral shape of NGC~5548 as used by
Mathur et al. (1995)\markcite{Mathur95}, which contains a soft X-ray excess.
The models are computed for constant density ($n_H = 10^3~\rm cm^{-3}$) clouds
in thermal equilibrium.
(For densities $< 10^{11}~\rm cm^{-3}$ there are no density-dependent effects
in our calculations or results.)
The models are described by their ionization parameter $U$
and their total column density $N$.

\begin{figure}[!t]
\plotfiddle{"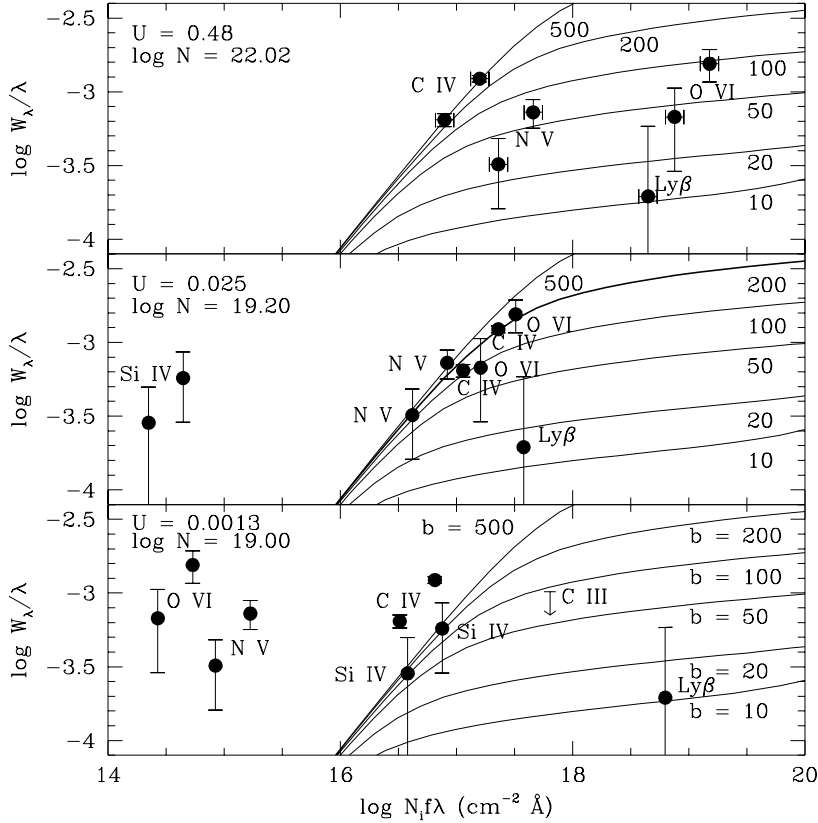"}{4.0in}{0}{60}{60}{-180}{-100}
\caption{
{\it Top Panel:} The observed EWs of the UV absorption lines in NGC~3516 are
plotted on curves of growth using column densities predicted by the single warm
absorber fit to the ASCA X-ray spectrum (Kriss et al. 1996).
This model has $U = 0.48$ and a total column density of
$10^{22.02}~\rm cm^{-2}$.
Points are plotted at a horizontal position determined by the column density
for the given ion in the model with a vertical coordinate determined by the
observed EW for the corresponding absorption line.
The vertical error bars are from Table 2, and the horizontal error bars are
the range in column density allowed by the uncertainty in the fit to the
ASCA spectrum.
The predicted column of Si~{\sc iv} in this model is too low to appear on the
plot.
The thin solid lines show predicted EWs as a function of column density
for Voigt profiles with
Doppler parameters of $b = 10$, 20, 50, 100, 200, and $500~\rm km~s^{-1}$.
A model that fits the data would have all points lying on one of these curves.
This model cannot simultaneously match both the UV and the X-ray absorption.
{\it Center Panel:} The observed EWs are plotted for column densities
predicted by a model with $U = 0.025$ and a total column density of
$10^{19.2}~\rm cm^{-2}$.
The heavy solid curve at $b = 200~\rm km~s^{-1}$ gives
a good match to the observed
EWs of the {\sc C~iv}, {\sc N~v}, and {\sc O~vi} doublets,
but the predicted EW of Si~{\sc iv} is far below the observed value.
{\it Bottom Panel:} The observed EWs are plotted for column densities
predicted by a model with $U = 0.0013$ and $N = 10^{19.0}~\rm cm^{-2}$.
$b = 20~\rm km~s^{-1}$ can match the observed EWs of
Si~{\sc iv} and Ly$\beta$ and satisfy the upper limit on the EW of
{\sc C~iii} $\lambda977$.
\label{n3516uv4pp.ps}
}
\end{figure}

Single-zone photoionization models of the absorbing gas in 3C~351
(\cite{Mathur94}) and in NGC~5548 (\cite{Mathur95}) can produce both the
X-ray and UV absorption seen in these two AGN.
We now test whether this is true for NGC~3516.
The single warm absorber fit to the ASCA spectrum of NGC~3516 has $U = 0.48$
and $\log N = 22.02$.
Given the column densities predicted by this model for the UV ions,
we can place the observed EWs on curves of growth (Figure \ref{n3516uv4pp.ps}).
If the model self-consistently accounts for both the X-ray and UV absorbers,
all the plotted EWs should lie on a single curve of growth.
As shown in the top panel of Figure \ref{n3516uv4pp.ps}, they do not.
The predicted column densities of neutral hydrogen and Si~{\sc iv} are orders
of magnitude below what we observe in the HUT spectrum, and a single
Doppler parameter does not give a good match to the remaining UV lines.
Choosing $b = 70~\rm km~s^{-1}$ fits the {\sc N~v} and {\sc O~vi} doublets
within the errors, but this underpredicts the strength of the {\sc C~iv}
doublet by factors of several.
This situation is similar to the difficulties noted by
Kriss et al. (1995)\markcite{Kriss95} in trying to account for the
low-ionization UV absorption lines
in NGC~4151 with a single warm absorber model.

The warm absorber models computed with the alternative ionizing continua
give similar results, with the predicted columns of the UV ions in each
model within tens of percent of each other.
This insensitivity to the precise shape of the ionizing spectrum is
not surprising.  For broad ionizing continua
such as power laws or broken power laws,
even the earliest photoionization calculations (e.g., \cite{TTS69})
showed that models were much more sensitive to ionization parameter
than to spectral shape.
In Table \ref{tab-columns} we compare
the columns inferred from the UV absorption lines
assuming they are optically thin, as indicated by their doublet ratios,
to the column densities predicted by the various models.
The similarity of the models to each other is apparent, as is the
large disagreement with the observations, particularly for Si~{\sc iv}.

\begin{deluxetable}{ l c c c c}
\tablecaption{Observed and Predicted Absorption Columns in NGC 3516
\label{tab-columns}}
\tablecolumns{5}
\tablewidth{6.5in}
\tablehead{
\colhead{Ion} & \colhead{$N_{obs}$\tablenotemark{a}} & \colhead{$N_{Model
1}$\tablenotemark{b}} & \colhead{$N_{Model 2}$\tablenotemark{c}} &
\colhead{$N_{Model 3}$\tablenotemark{d}}\\
\colhead{} & \colhead{($\rm cm^{-2}$)} & \colhead{($\rm cm^{-2}$)} &
\colhead{($\rm cm^{-2}$)} & \colhead{($\rm cm^{-2}$)}}
\startdata
{\sc H i}   & $3.5 \times 10^{17}$ & $5.0 \times 10^{16}$ & $1.7 \times
10^{17}$ & $1.8 \times 10^{16}$ \nl
{\sc C iv}  & $4.7 \times 10^{14}$ & $4.7 \times 10^{14}$ & $6.0 \times
10^{14}$ & $5.4 \times 10^{14}$ \nl
{\sc N v}   & $4.1 \times 10^{14}$ & $2.1 \times 10^{15}$ & $2.5 \times
10^{15}$ & $3.2 \times 10^{15}$ \nl
{\sc O vi}  & $1.3 \times 10^{15}$ & $9.8 \times 10^{16}$ & $1.1 \times
10^{17}$ & $1.4 \times 10^{17}$ \nl
Si {\sc iv} & $1.1 \times 10^{14}$ & $< 1 \times 10^{10}$ & $< 1 \times
10^{10}$ & $< 1 \times 10^{10}$ \nl
\enddata
\tablenotetext{a}{Column density assuming the observed absorption lines
are optically thin.  The {\sc H~i} column is from the Lyman-limit fit.}
\tablenotetext{b}{Predicted column density for the single-zone warm absorber
model with the NGC~3516 ionizing spectrum, $U = 0.48$ and log $N$ = 22.02.}
\tablenotetext{c}{Predicted column density for the warm absorber model with
the $\nu^{-0.78}$ ionizing spectrum, $U = 0.16$ and log $N$ = 22.02.}
\tablenotetext{d}{Predicted column density for the warm absorber model with
the NGC~5548 ionizing spectrum, $U = 1.47$ and log $N$ = 22.02.}
\end{deluxetable}

The two-component warm absorber fit favored by the ASCA
data also fails as an adequate description of the UV absorption.
The lower ionization zone with $U = 0.32$ and $\log N = 21.84$
still falls short in matching the
Si~{\sc iv} and {\sc H~i} by orders of magnitude, and once again a single
Doppler parameter fails to match the observations of the remaining UV lines.
The only UV-absorbing species with a significant column at any Doppler
parameter in the higher ionization zone ($U = 1.66$, $\log N = 22.15$)
is {\sc O~vi}.

Producing the observed EWs of the UV lines requires regions of lower ionization
and lower column density than those producing the X-ray absorption.
For $U = 0.025$, $\log N = 19.2~\rm cm^{-2}$,
and $b = 200~\rm km~s^{-1}$ we find a good match to the observed
EWs of the {\sc C~iv}, {\sc N~v}, and {\sc O~vi} doublets
as shown in the center panel of Figure \ref{n3516uv4pp.ps}.
The predicted EW of the Ly$\beta$ line is $\sim 2 \sigma$ higher than observed,
but enhancing the overall metal abundance by a factor of a few can correct
this.
However, the predicted EW of
Si~{\sc iv} in this model is orders of magnitude below the observed value.
Similar problems are encountered in matching the inferred column densities of
Si~{\sc iv} absorption in models of broad-absorption line (BAL) QSOs
(\cite{WTC}).
{\it Ad hoc} adjustments to the ionizing continuum can
alleviate the disagreement to some extent (\cite{WTC}), but even the
most extreme changes do not bring the predictions within the bounds of the
observations, again showing the relative insensitivity of photoionization
models to the shape of the ionizing spectrum.
Producing enough Si~{\sc iv} absorption requires
a zone of even lower ionization.
Choosing $U = 0.0013$ and $\log N = 19.0$ matches the Si~{\sc iv} EWs without
contributing much to the columns of the higher ionization lines.
This model is illustrated in the lower panel of Figure \ref{n3516uv4pp.ps}.
Limits on the EWs of lower ionization species such as Ly$\beta$ and
{\sc C~iii} $\lambda 977$, however,
require $b < 50~\rm km~s^{-1}$.
This lower ionization zone with low Doppler parameter is potentially the
origin of the optically thick Lyman limit system, but the predicted
column of neutral hydrogen in this model is only
$N_{H I} = 7.7 \times 10^{16}~\rm cm^{-2}$, a factor of 4 below the observed
value.

The multiple zones considered above span a range of $10^3$ in both
column density and ionization parameter, and
they form a nearly orthogonal set
that makes it possible to consistently model the UV and X-ray absorption
with separate regions of gas in very different physical states.
For Doppler parameters at the sound speed
of the two high ionization zones producing the X-ray absorption
($30~\rm km~s^{-1}$ in the lower ionization zone,
$70~\rm km~s^{-1}$ in the higher),
the predicted contribution of the X-ray
absorbing gas to the EWs of the {\sc C~iv}, {\sc N~v}, and {\sc O~vi} lines
is only 20--30\% of their observed values.
The UV zones producing the {\sc C~iv}, {\sc N~v}, {\sc O~vi} and Si~{\sc iv}
lines have total columns with negligible impact on the X-ray opacity.
Since the X-ray absorbing gas is predicted to have a significantly lower
Doppler parameter than the gas producing the bulk of the UV absorption lines,
high resolution spectra in the UV should be able to identify
kinematic components associated with the different zones.

Several previous authors have suggested that the UV absorption lines and
the X-ray absorption seen in some AGN may have a common origin
in BELR clouds (\cite{FM82}; \cite{RMH86}).
However, Mathur et al. (1994)\markcite{Mathur94} show that the physical
conditions of the UV and X-ray absorber in 3C~351 are different from those of
the BELR clouds, with the X-ray warm absorber requiring a much higher
ionization
parameter than the clouds producing the broad emission lines.
For NGC~3516, Voit, Shull, \& Begelman (1987)\markcite{Voit87}
argued that the observed absorption profile (the {\sc C~iv} absorption in
the 1980's was much broader and stronger than observed here)
required 10--30 BELR clouds along the line of sight and that photoionized
{\sc C~iv} could not be present in the outer clouds due to shielding by
the inner ones.
The lack of observed Mg~{\sc ii} absorption was another difficulty for
models involving BELR clouds.
They concluded that the absorption must arise in optically thin,
outflowing material with a density exceeding $10^5~\rm cm^{-3}$, given
the observed variability, and that it could not be associated with
typical BELR clouds.

Our conclusions are similar, even though the absorption is now much weaker.
The ionization parameter we derive for the broad emission-line clouds
($U \sim 0.035$) is an order of magnitude lower than that required by the
X-ray absorbing gas, making it highly unlikely that the X-ray absorption
occurs in BELR clouds.
$U \sim 0.035$, however, is similar to the ionization parameter we find for the
zone producing the {\sc C~iv}, {\sc N~v},
and {\sc O~vi} absorption lines ($U \sim 0.025$).
Although these absorption lines and the broad emission lines share
comparable ionization parameters,
the total columns of the BELR clouds are $\sim 10^3 \times$ higher, as the
broad emission line clouds are optically thick in these transitions,
but the absorbing clouds are optically thin.
This column density constraint is a poor one, however, because the BELR is
highly stratified, and there may well be a population of clouds that both
make the absorption lines and contribute to the broad-line emission.
The velocity distribution of such clouds would have to be rather peculiar,
however, to project a velocity width of only $200~\rm km~s^{-1}$ along
the line of sight from an ensemble with an emission line width of
$\sim 4000~\rm km~s^{-1}$.
We conclude that we still cannot unambiguously establish the location of the
absorbing gas,
but it must be physically distinct from the broad emission line clouds.

Although we have described the UV and X-ray absorbing zones as discrete
entities, it is quite likely that there is a broad, possibly continuous,
distribution of parameters such as one might find in an outflowing wind,
either from the accretion disk or from the surface of the obscuring torus.
As discussed in the introduction, NGC~3516 is one of the rare Seyfert 1
galaxies that shows an extended, biconical NLR
(see Golev et al. 1995\markcite{Golev95} for the best images).
Among Seyfert 1's, only NGC~4151 has such an extensive NLR with a biconical
morphology (\cite{Pogge89}; \cite{Evans94}; \cite{SK96}).
The biconical morphology in NGC~3516 implies that our line of sight passes
close to the surface of the obscuring torus as suggested for NGC~4151 by
Evans et al. (1993).\markcite{Evans93}
The opaque Lyman limit is another characteristic that NGC~3516 shares
with NGC~4151.
These features are unusual in combination for a Seyfert 1, and
their presence in yet another galaxy strengthens the case for
collimation of the ionizing radiation by the absorbing gas.

The complexity of the UV and X-ray absorption that both NGC~3516 and NGC~4151
exhibit suggest that the inclination of the source relative to the observer
may lead to the differences between these two objects and
the apparently simpler cases considered by
Mathur et al. (1994,1995).\markcite{Mathur94}\markcite{Mathur95}
The simpler warm absorbers with only high-ionization UV absorption lines may be
viewed at higher inclination,
further away from the denser medium near the torus.
Such geometrical differences may ultimately help us to understand the location
and origin of the warm absorbing gas in AGN.

\acknowledgments
This work was supported by NASA contract NAS 5-27000 to the Johns Hopkins
University.

\vfill\eject

\end{document}